\documentclass{nature}

\usepackage{caption}
\usepackage{float}
\usepackage{graphicx}
\usepackage[scaled]{helvet}
\usepackage{amsmath}
\usepackage{amssymb}
\usepackage{bm}
\usepackage{xfrac}
\newcommand*{\myfont}{\fontfamily{phv}\selectfont}
\usepackage[pdfstartpage=1]{hyperref}
\usepackage{anyfontsize}
\usepackage{ragged2e}
\usepackage{color}

\definecolor{CLBlue}{rgb}{0, .25, .8}
\definecolor{MyBlue}{rgb}{0, .24, .40}
\definecolor{MyTurquoise}{rgb}{0, .53, .49}
\definecolor{MyGreen}{rgb}{0, .35, 0}
\definecolor{MyOrange}{rgb}{.8, .46, 0}
\definecolor{MyRed}{rgb}{.57, .07, 0}
\definecolor{MyPurple}{rgb}{.46, .1, .46}

\title{Quantifying the cost of network computations to unpack structure--function relationships in the brain.}

\author{Suman S. Kulkarni$^{1}$, Jason Z. Kim$^{2}$, Panagiotis Fotiadis$^{3}$, Fabio Pasqualetti$^{4}$ \& Dani S. Bassett$^{5, 6,*}$}

\begin{document}
\maketitle

\begin{affiliations}
\item Department of Physics \& Astronomy, University of Pennsylvania, PA 19104, USA
\item Department of Physics, Cornell University, Ithaca, NY 14853, USA
\item Department of Neurology, Massachusetts General Hospital, Harvard Medical School, Boston, MA 02114, USA
\item Department of Electrical Engineering and Computer Science, University of California, Irvine, CA 92697, USA
\item Departments of Biomedical Engineering and Psychology, Yale University, New Haven, CT 06520, USA
\item Wu Tsai Institute, Yale University, New Haven, CT 06520, USA
\item[*] Corresponding author: dani.bassett@yale.edu
\end{affiliations}

\noindent {\large \myfont \textbf{Abstract}}
\vspace{-28pt}

\noindent\rule{\textwidth}{.5pt}

The brain supports computations through coordinated patterns of activity on an underlying network. These networks---from microscale navigational circuits in insects to macroscale brain areas in humans---are organized in structured ways that are thought to support their function. We seek a unifying quantitative framework to understand how network structure shapes the computations a network can readily support. To do so, we frame computation as a goal-directed transition of activity and quantify its cost on a given network using control theory. We then define the distribution of costs across all possible transitions as a \emph{computational affordance landscape} that encodes which computations a network structure readily supports. We apply this framework to a circuit model for how insects maintain a sense of direction and show that updating orientation is the least costly computation, with predicted inputs consistent with known circuitry. In the human brain, we find that the affordance landscape varies systematically with the functional role of each network. Sensory networks display more heterogeneous landscapes (reflecting their role in specialized information processing), whereas association networks display more homogeneous landscapes (reflecting their role in generalized information processing). In recurrent neural networks trained on cognitive tasks, we show that learning progressively increases landscape heterogeneity, reshaping the distribution of affordable computations. Generally, we establish a quantitative framework for studying relationships between structure and computation in neural circuits, with future applications extending to other biological and physical networks.

\noindent {\large \myfont \textbf{Main}}
\vspace{-28pt}

\noindent\rule{\textwidth}{.5pt}
\noindent 

In the brain, coordinated patterns of activity across neurons and brain regions give rise to a remarkable range of functions, from navigation and decision-making to learning and memory~\cite{bassett2011dynamic,rugg2013brain,bressler2010large,xie2002double,park2013structural}. These patterns emerge from networks of interacting neurons or brain regions that span scales ranging from individual synaptic connections to large-scale cortical circuits~\cite{kulkarni2025toward,park2013structural,bassett2017network}. Each network receives inputs from other brain areas that actively drive it through sequences of activity states, implementing computations that underlie neural function~\cite{vyas2020computation,shenoy2013cortical,turner2017angular,semedo2019cortical}. Extensive efforts have characterized the anatomical structure of interactions in the brain, from synaptic-level connectomes~\cite{lin2024network,varshney2011structural} to white-matter tracts~\cite{van2013wu,bullmore2009complex}. Parallel efforts have also investigated the statistical structure of neural activity, uncovering functional correlations in activity across scales~\cite{power2011functional,hipp2012large,schneidman2006weak,stringer2019high} and low-dimensional structure in population activity~\cite{jazayeri2021interpreting,mehrkanoon2014low,gallego2017neural,cunningham2014dimensionality}. Together, these efforts have begun to reveal how structure and activity are related~\cite{bettinardi2017structure,park2013structural,hermundstad2014structurally,wang2015understanding}, yet several basic questions remain ~\cite{honey2010can,fotiadis2024structure}. For a given network, which computations does its structure naturally support? And what does this relation reveal about biological function?

Addressing these questions requires a general framework that maps network structure directly onto a quantitative measure of how readily any given pattern of activity can be obtained---a notion of \emph{computational affordance} given a network structure. Control theory provides tools to develop this notion
~\cite{bechhoefer2021control,kim2020linear,gu2015controllability}. We define a computation as a controlled transition of network activity towards a functionally-relevant state; in this setting, a network exists in an initial state, receives inputs from other regions, and is driven to a novel, functionally-relevant state. We quantify the cost of a computation in terms of the inputs required to implement it. Networks whose structure intrinsically favors a computation can implement it at low cost, while those whose structure is misaligned require a higher cost. Building on classical tools from control theory~\cite{parkes2024network,gu2015controllability,kim2018role}, we define an affordance landscape that captures the full distribution of costs across all possible activity transformations, and use it to relate network structure directly to biological function.

We develop this framework and apply it to neural circuits at multiple scales, with the goal of understanding structure--computation relationships. We first show how the cost of any computation can be expressed in terms of network structure; specifically, it depends on how the computation aligns with the affordance landscape. We then apply the framework to a model of the head direction circuit in \emph{Drosophila melanogaster}, a system where structure, dynamics, and computation are well characterized~\cite{seelig2015neural,noorman2024maintaining}. We find that the network's energetically cheapest computation, informed by its structure, corresponds precisely to its known biological function of updating the heading representation. We also recover the structure of inputs required to implement this computation, finding them consistent with the known anatomy of the fly head direction circuit~\cite{turner2020neuroanatomical,green2017neural}. Turning to the human brain, we find that the heterogeneity of the affordance landscape varies systematically along the principal sensorimotor-association axis of the cortex~\cite{sydnor2021neurodevelopment}. Sensory networks display more heterogeneous landscapes, consistent with their role in unimodal processing, whereas association networks display more homogeneous landscapes, consistent with their role in flexible, domain-general processing~\cite{margulies2016situating}. Finally, we show that training artificial recurrent neural networks on specific tasks progressively increases the heterogeneity of their affordance landscapes, suggesting that goal-directed learning sculpts the landscape of affordable computations. Together, these results establish a quantitative framework for understanding how network structure shapes computational function across scales.

\noindent {\large \myfont Quantifying the cost of network-level computations}

\begin{figure}[h!]
{   \centering
    \includegraphics[width=0.95\linewidth]{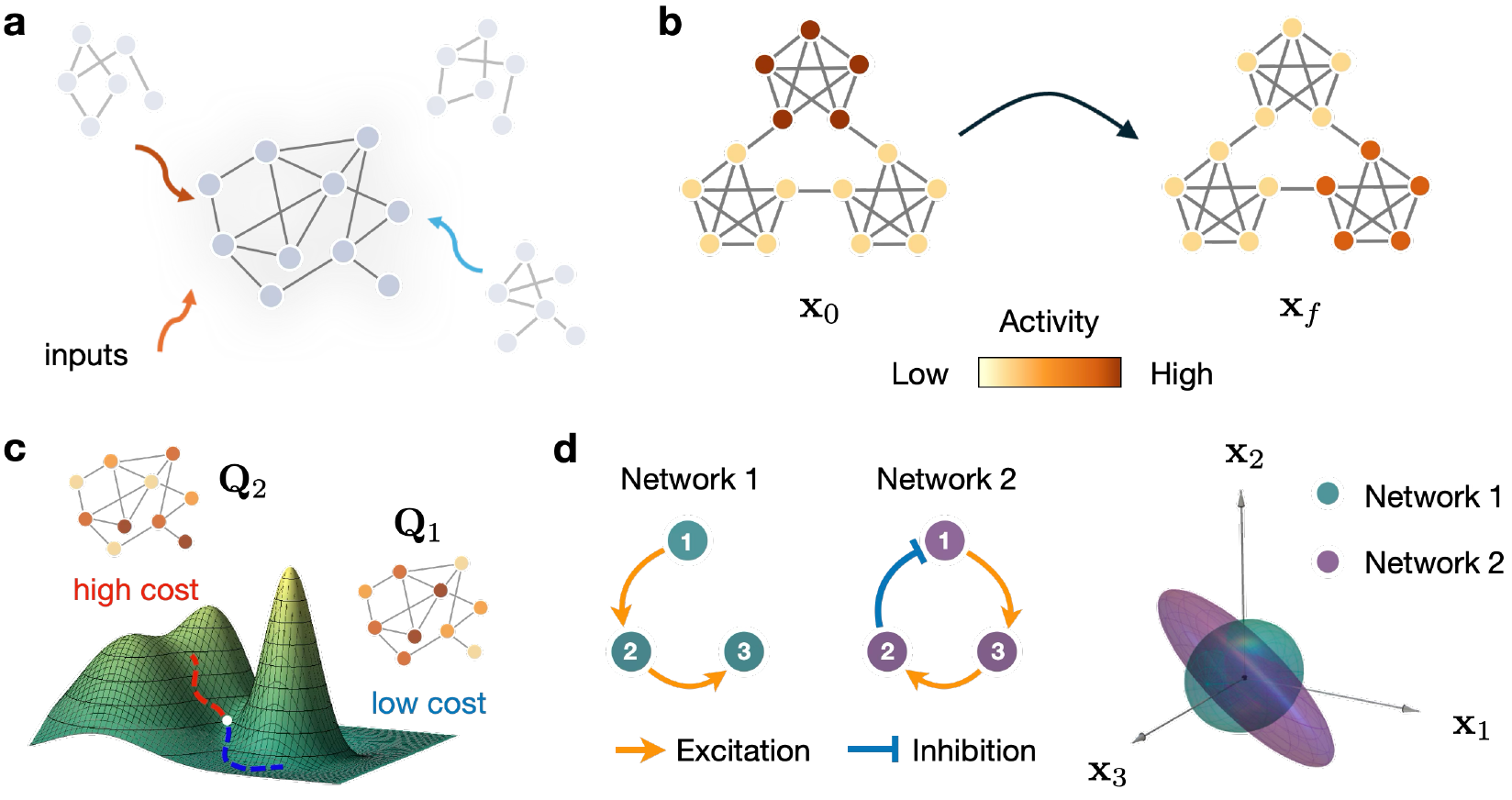} \\
    \raggedright
    \captionsetup{labelformat=empty}
    {\spacing{1.25} \caption{\small \textbf{Fig.~\ref{fig:Fig1} $|$ Network structure shapes the cost of computations.} \textbf{a}, A network in one brain area receives inputs from other brain areas, driving transitions between states of activity. \textbf{b}, A computation is a goal-directed transition of network activity from an initial state $\vec{\mathbf{x}}_0$ to a functionally relevant target state $\vec{\mathbf{x}}_f$. \textbf{c}, The cost of reaching a target state varies across activity space, reflecting an underlying computational affordance landscape that is shaped by network structure. Transitions to certain patterns of activity are low in cost to drive (for example, a transition to pattern ${\mathbf{Q}}_1$), while others are high in cost to drive (for example, a transition to pattern ${\mathbf{Q}}_2$). \textbf{d}, Two networks of 3 nodes with distinct network structure (left). The set of activity states reachable within a unit input cost forms an ellipsoid in state space (right), with principal axes aligned to the eigenvectors of the Gramian $\mathcal{W}$ [Eq. \eqref{eq:min_cost_gramian}] and radii equal to the square roots of its eigenvalues. \label{fig:Fig1}}}
}
\end{figure}

We consider a network of $N$ interacting nodes, which we conceptualize as representing brain regions or neurons, and we associate a real value (state) with each node that describes its activity. The set of all nodes' states at time $t$ forms a vector $\mathbf{x}(t) \in \mathbb{R}^N$, which we refer to as the state of the network. The network supports computations through $\mathbf{x}(t)$, which evolves according to a function $f$ that captures the network's internal interactions and external drive $\mathbf{u}(t) \in \mathbb{R}^m$ (Fig.~\ref{fig:Fig1}\,\textbf{a}) such that
\begin{equation}\label{eq:full_dynamics}
    \frac{d \, \mathbf{x}(t)}{dt} = f(t, \mathbf{x}(t) \,, \mathbf{u}(t)) \,.
\end{equation}

\noindent We define a computation as a goal-directed transition of network activity from an initial state $\mathbf{x}_0$ to a functionally relevant target state $\mathbf{x}_f$, driven by $\mathbf{u}(t)$ over a time horizon $T$ (Fig.~\ref{fig:Fig1}\,\textbf{b}). We quantify the cost of a computation by the energy of the input required to implement it,
\begin{equation}
    \mathcal{C} = \int_0^T \|\mathbf{u}(t)\|^2 \, dt \,.
\end{equation}
For analytical tractability, we linearize the dynamics in \eqref{eq:full_dynamics} about an equilibrium point ($\mathbf{x}^*$, $\mathbf{u}^*$), and we express the state as $\mathbf{x}(t) = \mathbf{x}^* \,+\, \mathbf{r}(t)$ for small deviations $\mathbf{r}(t)$ about equilibrium. In particular, the initial and target states correspond to $\mathbf{r}_0 = \mathbf{x}_0 - \mathbf{x}^*$ and $\mathbf{r}_f = \mathbf{x}_f - \mathbf{x}^*$, respectively. To first order, the deviations evolve as $\dot{\mathbf{r}}(t) = W \mathbf{r}(t) \,+\, B \, \mathbf{v}(t)$, where $W \in \mathbb{R}^{N \times N}$ is the Jacobian of $f$ at $\mathbf{x}^*$, $B \in \mathbb{R}^{N \times m}$ specifies how the inputs project onto the nodes, and $\mathbf{v}(t) = \mathbf{u}(t) \,-\, \mathbf{u}^*$ describes the deviation of the input from equilibrium. Under this approximation, the minimum cost of driving the network to the target functional state has a closed form (Methods). Intuitively, the input must compensate for any discrepancy $\mathbf{D} = \mathbf{r}_f \, - \, e^{WT} \mathbf{r}_0$ between the target state $\mathbf{r}_f$ and the homogeneous solution $e^{WT} \mathbf{r}_0$ (i.e., the state of the network at time $T$ when $\mathbf{v}(t) = 0$). The minimum cost at which this computation can be achieved is
\begin{equation}\label{eq:min_cost_gramian}
    \mathcal{C}^{\mathrm{min}} = \mathbf{D}^{\top} \, \mathcal{W}^{-1} \, \mathbf{D}, \quad \textrm{where} \quad \mathcal{W} = \int_0^T e^{Ws} B B^{\top} e^{W^{\top}s} \, ds
\end{equation}
is the controllability Gramian of the network~\cite{bechhoefer2021control,kim2018role,kim2020linear} (Methods). The Gramian encodes how the network interactions $W$ and its input structure $B$ jointly determine which activity patterns are low in cost (i.e., easy) to produce and which activity patterns are high in cost (i.e., hard) to produce. The inverse Gramian $\mathcal{W}^{-1}$ hence defines a \emph{computational affordance landscape} over state space that is shaped by network and input structure (See schematic in Fig.~\ref{fig:Fig1}\,\textbf{c}). 

The dependence on network structure is also directly visible in the set of states reachable within unit cost. For linearized dynamics, these states form an ellipsoid in $\mathbb{R}^{N}$ with principal axes aligned to the eigenvectors of $\mathcal{W}$ and radii equal to the square roots of its eigenvalues. To build intuition, we consider two networks of $N = 3$ nodes with distinct structure, assuming each node receives input independently ($B = \mathbb{I}$), as shown in Fig.~\ref{fig:Fig1}\,\textbf{d}. The first network is a feedforward chain in which node 1 excites node 2, which in turn excites node 3. The second network is a feedforward chain in the opposite direction (i.e., node 1 excites node 3, which in turn excites node 2), with an inhibitory connection from node 2 to node 1. The two networks result in strikingly different ellipsoids of unit cost (Fig.~\ref{fig:Fig1}\,\textbf{d}, right panel) that differ in both orientation and elongation. In particular, we note that in the second network, the inhibitory interaction from node 2 to 1 causes the network to favor an activity pattern of node 1 that is anti-correlated with that of node 2. Hence, the resulting ellipsoid shape. In summary, the network structure directly shapes the cost of computations and determines which computations a network supports efficiently. We next quantify this dependence explicitly for a specific network structure and target computation.

\noindent {\large \myfont Cost of computations reflects alignment with computational affordance landscape}

We consider a computation that is a simple superposition, in which a network is driven from a baseline state to a modulated state, offset by a pattern $\mathbf{Q}$. A natural example of such a superposition is neuromodulation of large cortical circuits by the \emph{locus coeruleus}, which broadcasts signals that shift the activity of these neural populations~\cite{noei2022distinct,breton2021locus}. Assuming that the baseline activity state is a stable fixed point, the cost of this modulation is $\mathcal{C}^{\mathrm{min}}(\mathbf{Q}) = \mathbf{Q}^{\top} \, \mathcal{W}^{-1} \, \mathbf{Q}$. To understand what determines this cost, we expand $\mathbf{Q}$ in the eigenbasis of $\mathcal{W}$. Let $\{(w_i, \mathbf{v}_i)\}_{i = 1}^{N}$ denote the eigenvalue and eigenvector pairs of $\mathcal{W}$, where $\mathcal{W} \, \mathbf{v}_i = w_i \, \mathbf{v}_i$. We decompose the cost in the eigenbasis of $\mathcal{W}$ to obtain
\begin{equation}\label{eq:min_cost_eigen}
   \mathcal{C}^{\mathrm{min}}(\mathbf{Q}) = \sum_i \frac{c_i^2}{w_i}, \quad \text{where} \quad c_i = \mathbf{v}_i^\top \, \mathbf{Q} \,.
\end{equation}
Each term in this sum reflects how strongly the target pattern $\mathbf{Q}$ projects onto a mode $\mathbf{v}_i$ of $\mathcal{W}$ and is weighted by $w_i^{-1}$, the eigenvalue of the affordance landscape along that mode. In other words, $w_i^{-1}$ is the cost of exciting mode $\mathbf{v}_i$ by a unit amount.

\begin{figure}[t]
    \centering
    \includegraphics[width=0.85\linewidth]{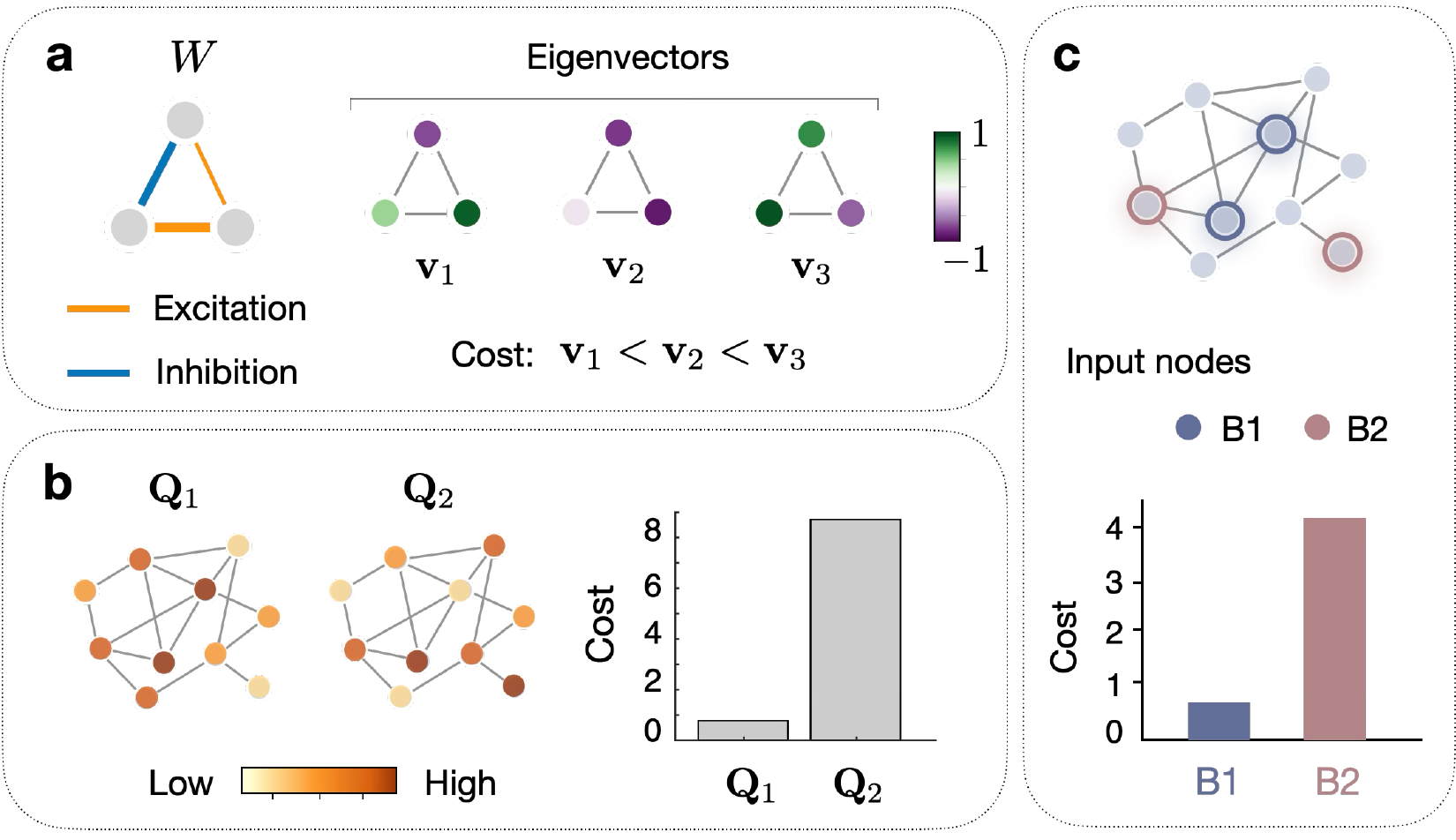} \\
    \raggedright
    \captionsetup{labelformat=empty}
    {\spacing{1.25} \caption{\small \textbf{Fig.~\ref{fig:Fig2} $|$ The cost of a computation reflects its alignment with the computational affordance landscape.} \textbf{a}, We consider a small network with excitatory (orange) and inhibitory (blue) interactions, where every node can receive input. Any target computation can be decomposed in terms of the network's eigenmodes $\mathbf{v}_i$, and the cost of driving each mode is determined by its eigenvalue [Eq.~\eqref{eq:min_cost_eigen}]. The cost of a computation can therefore be expressed in terms of network structure. \textbf{b}, The cost of driving a target pattern depends on its alignment with the affordance landscape. Pattern $\mathbf{Q}_1$ aligns with modes of the landscape that are low in cost, while pattern $\mathbf{Q}_2$ does not and is therefore higher in cost. \textbf{c}, More generally, the cost of a computation also depends on which nodes receive input. For a fixed network and target pattern, driving the same computation is low in cost under input structure $B_1$ but substantially higher in cost under $B_2$. Together, network and input structure jointly shape the affordance landscape and the cost of computations. \label{fig:Fig2}}}
\end{figure}

The dependence on network structure simplifies considerably if each node receives input independently $(B = \mathbb{I})$ and if the interaction matrix $W$ is symmetric (i.e., $W = W^{\top}$). In this setting, $\mathcal{W}$ shares the same eigenbasis as $W$, and the cost of each mode takes the form
\begin{equation} \label{eq:w_i_inverse_simplified}
    w_i^{-1}  = \frac{2 \lambda_i}{e^{2\lambda_i T} - 1}, 
\end{equation}
where $\lambda_i$ are the eigenvalues of $W$. We observe that the cost~\eqref{eq:w_i_inverse_simplified} is a monotonically decreasing function of $\lambda_i$. For stable $W$ (i.e., $\lambda_i < 0$ for all $i$), we interpret the cost of each mode in terms of its timescale $\tau_i \, = \, 1/|\lambda_i|$: slower timescale modes ($\lambda_i \approx 0$) are easy to excite, while faster timescale modes ($\lambda_i \ll 0$) are difficult to excite. 

We demonstrate this relationship between cost and network structure for a small, symmetric network with a mix of excitatory and inhibitory connections (Fig.~\ref{fig:Fig2}\,\textbf{a}, left). The network's eigenmodes $\mathbf{v}_1, \mathbf{v}_2, \mathbf{v}_3$ form a natural basis for any target pattern (Fig.~\ref{fig:Fig2}\,\textbf{a}, right), and the cost of driving the network along each mode follows directly from its eigenvalue as per Eq.~\eqref{eq:w_i_inverse_simplified}. For this network, the modes are ordered $\mathbf{v}_1 < \mathbf{v}_2 < \mathbf{v}_3$ in cost. The cost of any target pattern $\mathbf{Q}$ then depends on how it projects onto this basis [See Eq.~\eqref{eq:min_cost_eigen}]. To illustrate this, we consider a larger network in Fig.~\ref{fig:Fig2}\,\textbf{b}, and we compare the cost of two target modulations $\mathbf{Q}_1$ and $\mathbf{Q}_2$. Pattern $\mathbf{Q}_1$ aligns with low-cost modes (which predominantly involve co-activation of the hub nodes of the network) and is therefore lower in cost. Pattern $\mathbf{Q}_2$ aligns with expensive modes and is therefore higher in cost (Fig.~\ref{fig:Fig2}\,\textbf{b}, right). 

The cost of a computation depends not only on the network structure and target pattern but also on which nodes receive input [Eq.~\eqref{eq:min_cost_gramian}]. For a fixed network and target pattern, different input matrices $B$ induce distinct affordance landscapes. For example, in Fig.~\ref{fig:Fig2}\,\textbf{c}, we compare the cost of driving the same target pattern $\mathbf{Q}_1$ (from Fig.~\ref{fig:Fig2}\,\textbf{b}) under two distinct input structures, $B_1$ and $B_2$. The landscape induced by $B_1$ has low-cost modes that align closely with $\mathbf{Q}_1$, making the computation low in cost; the landscape induced by $B_2$ does not, making the same computation high in cost.

Taken together, the cost of any computation depends on how it aligns with the computational affordance landscape, which is in turn shaped by network and input structure. In Supplementary Information Sec.~\ref{sec:oscillations}, we demonstrate how our notion of computation also applies to oscillatory computations, illustrating how neural circuits can act as filters~\cite{ito2007frequency,campbell2025hardwired} (Fig.S~\ref{fig:Fig6}\,\textbf{a}) and can selectively route information~\cite{colgin2009frequency, khamechian2019routing} (Fig.S~\ref{fig:Fig6}\,\textbf{b}).

\noindent {\large \myfont Structure, computation and function in a circuit for navigation}

\begin{figure}[t!]
    \centering
    \includegraphics[width=0.8\linewidth]{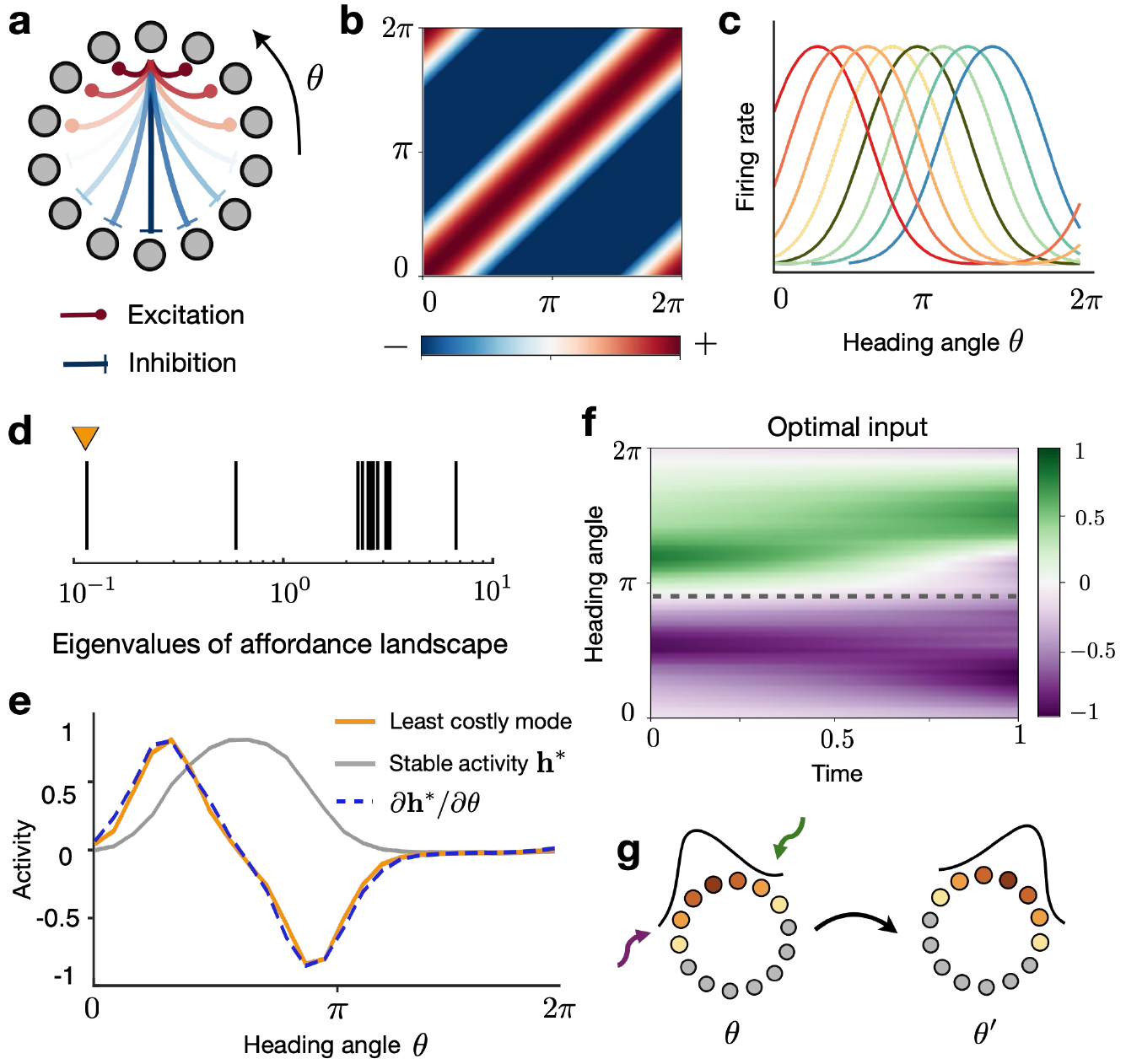} \\
    \raggedright
    \captionsetup{labelformat=empty}
    {\spacing{1.25} \caption{\small \textbf{Fig.~\ref{fig:Fig3} $|$ Structure, function and computation in a circuit for navigation.} \textbf{a}, Schematic of a ring-attractor network, which maintains a representation of the head direction $\theta$, via local excitation (red) and broad inhibition (blue). \textbf{b}, Weight matrix of the synaptic weights in the ring-attractor network. \textbf{c}, Localized bumps of activity that are stable at any heading orientation $\theta$. \textbf{d}, Eigenvalues of the computational affordance landscape evaluated about a stable bump of activity. The orange arrow indicates the lowest eigenvalue, which corresponds to the least costly mode in the affordance landscape. \textbf{e}, Stable bump of activity $\mathbf{h}^*$, its derivative with respect to heading $\partial \, \mathbf{h}^*/ \partial \theta$, and the least costly mode of the affordance landscape. \textbf{f}, Optimal input required to rotate the stable bump clockwise by 5 nodes (i.e., from orientation $\theta$ to $\theta^{\prime} = \theta \,+ \, 5 \cdot  [2\pi/N]$) as a function of time (x axis) and preferred heading angle (y axis). \textbf{g}, Schematic of the optimal inputs required to rotate a stable bump of activity clockwise, consisting of an increase in input ahead of the bump (in the direction of rotation) and a decrease in input behind the bump (in the direction opposite the rotation). \label{fig:Fig3}}}
\end{figure}

To demonstrate the utility of our framework, we turn to a concrete biological example where there is a clear convergence between network structure, dynamics, and computation: circuits for navigation. Many animals rely on specialized neural circuits that maintain an internal representation of direction as they navigate through the world~\cite{seelig2015neural,burak2009accurate,finkelstein2015three}. In the fruit fly \emph{Drosophila melanogaster}, heading direction is encoded by a population of neurons in the central complex, known as the EPG or ``compass" neurons~\cite{seelig2015neural,hulse2021connectome}. These neurons sustain a localized bump of activity that traverses a ring-like network as the fly turns, effectively functioning as a compass that tracks the animal's heading direction. Experimental and theoretical studies have shown that this circuit exhibits the structure and dynamics of a ring-attractor network~\cite{noorman2024maintaining,turner2020neuroanatomical,kim2017ring}, and we consider a recently proposed model that generates this representation of heading in small networks~\cite{noorman2024maintaining}.

The network consists of $N$ neurons with preferred orientations $\theta_j$ that uniformly tile the orientation space $[0, 2 \pi)$, with an angular separation of $\Delta \theta = 2\pi/N$ between consecutive neurons (Fig.~\ref{fig:Fig3}\,\textbf{a}). Localized and sustained activity is achieved through a recurrent connectivity by which neurons with similar tuning excite one another, and neurons with dissimilar tuning inhibit one another~\cite{noorman2024maintaining,khona2022attractor}. This connectivity is captured by the matrix $W_{jk} = J_I \, + \, J_E \, \cos(\theta_j \,-\, \theta_k)$, where $J_I$ and $J_E$ are the strengths of local excitation and global inhibition (Fig.~\ref{fig:Fig3}\,\textbf{b}). The total input activity to a neuron $h_j$ evolves according to
\begin{equation}\label{eq:ring_attractor_eq}
    \tau \frac{d h_j(t)}{dt} = - h_j(t) + \frac{1}{N} \sum_k W_{jk} \, \Phi(h_k) + I_j(t),
\end{equation}
where $\Phi(\cdot)$ is a nonlinear transfer function mapping the input to the firing rate, $I_j(t)$ is any external input to the neuron, and $\tau$ is a neural time constant. We take $\Phi(x) = [1 \, + \, \tanh (x)]/2$. For appropriate values of $(J_I, J_E)$, the network generates a continuum of localized bumps of activity that are stable at any heading angle $\theta \in [0, 2\pi)$~\cite{noorman2024maintaining}, as shown in Fig.~\ref{fig:Fig3}\,\textbf{c}.

We examine a ring attractor network with $N = 25$ neurons with a small amount of noise in the weight matrix (Methods). We compute the computational affordance landscape by linearizing Eq.~\eqref{eq:ring_attractor_eq} about a stable bump of activity $\mathbf{h}^*$ at orientation $\theta$ (see Supplementary Information Sec.~\ref{sec:SI_ring_attractor}). Strikingly, we observe a single mode in the landscape that is substantially lower in cost than all others (Fig.~\ref{fig:Fig3}\,\textbf{d}). This mode corresponds precisely to the derivative of the bump with respect to the heading angle (Fig.~\ref{fig:Fig3}\,\textbf{e}). That is, the network's cheapest computation, derived from our computational affordance landscape, is precisely its known function of updating the heading representation. In Supplementary Information Fig.\,S~\ref{fig:Fig7}, we also examine the second lowest-cost and highest-cost modes of the landscape for completeness.

Our framework also predicts the optimal structure of inputs required to update the heading direction. Using Eq.~\eqref{eq:optimal_input}, we find that the optimal input consists of an increase in input to neurons ahead of the activity peak (i.e., in the direction of rotation), and a decrease in input to neurons behind the activity peak (i.e., opposite to the direction of rotation), as shown in Fig.~\ref{fig:Fig3}\,\textbf{f}-\textbf{g}. This input profile aligns remarkably well with the known anatomy of the fly head direction system, in which two populations of shift neurons provide balanced input to the compass neurons at rest. When the fly turns and the bump requires an update, self-motion signals from the noduli break this balance, resulting in an increase in drive from one population (that increases the input ahead of the bump), and a decrease in drive from the other population (that decreases the input behind the bump) ~\cite{turner2020neuroanatomical,green2017neural}. 

Taken together, these results illustrate how our framework connects network structure and dynamics to computation in a way that yields direct insight into circuit function.

\noindent{\large \myfont Affordance landscape heterogeneity tracks the sensorimotor-association axis in the human brain }

\begin{figure}[h!]
    \centering
    \includegraphics[width=0.8\linewidth]{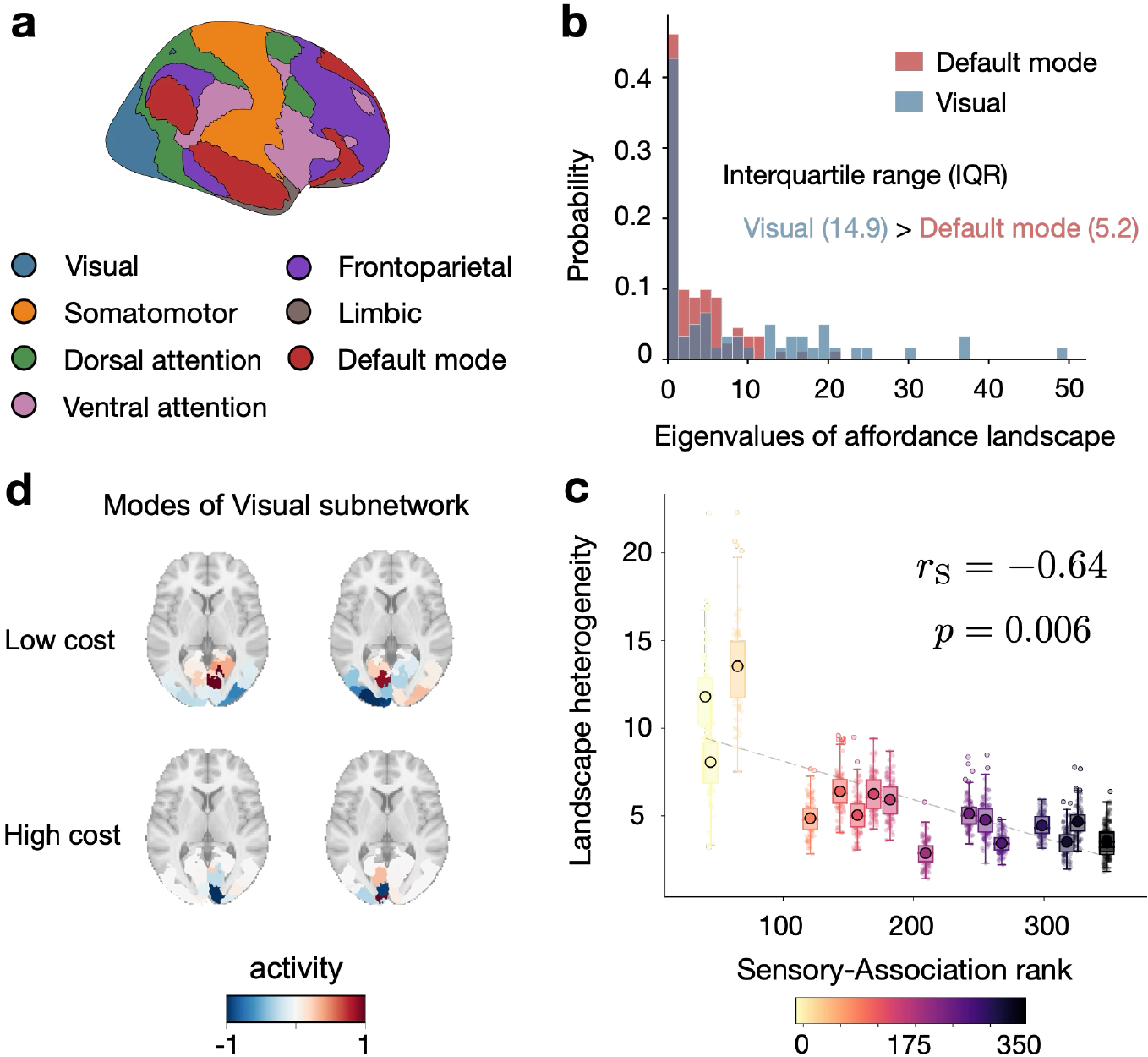} \\
    \raggedright
    \captionsetup{labelformat=empty}
    {\spacing{1.25} \caption{\small \textbf{Fig.~\ref{fig:Fig4} $|$ Affordance landscape heterogeneity tracks the sensorimotor-association axis in the human brain.} \textbf{a}, Cortical parcellation of the human brain into the 7 canonical networks. \textbf{b}, Distribution of eigenvalues of the affordance landscape for the visual and default mode networks in a representative individual. We report the interquartile range (IQR) of the affordance landscape eigenvalues for the two networks. \textbf{c}, Heterogeneity of the affordance landscape (IQR of the eigenvalues) for each of the 17 Yeo-atlas networks (y axis) plotted against each network's position along the sensorimotor-association (S-A) axis. Each point represents an individual subject, and box plots indicate the median and interquartile range across subjects. We report the Spearman correlation between the landscape heterogeneity and the S-A axis rank ($r_S = -0.64$, $p = 0.006$). \textbf{d}, Spatial distribution of the least expensive (top) and most expensive (bottom) modes of the affordance landscape for the visual network in a representative individual.  \label{fig:Fig4}}}
\end{figure}

We now turn to studying how network structure shapes computations in the human brain. The human cortex is organized into networks consisting of brain regions primarily implicated in specific functions, such as the visual network (processing visual information~\cite{wandell2007visual}) and the default mode network (involved in higher-order functions, such as self-reflection and future planning~\cite{buckner2008brain}). We illustrate the subdivision into 7 canonical networks according to the Yeo atlas~\cite{yeo2011organization} in Fig.~\ref{fig:Fig4}\,\textbf{a}. Each of these networks has distinct structural connectivity, raising a key question: how do these structural distinctions result in differences in the computations afforded? We hypothesize that differences in structure across networks produce computational affordance landscapes that reflect the functional specialization of each network.

To address this question, we examine structural connectivity from 100 independent subjects of the Human Connectome Project (HCP) and compute the affordance landscape for each network. The landscapes differ markedly across networks. For example, the visual network exhibits a substantially heavier tail in the distribution of costs than the default mode network (Fig.~\ref{fig:Fig4},\textbf{b}), indicating that some modes are much more expensive than others, channeling the network toward a subset of low-cost modes and suggesting specialization for a specific set of computations. By contrast, the default mode network has a more homogeneous landscape in which costs are similar across modes, enabling more uniform access to a broader range of activity patterns and suggesting generalization across computations. To quantify this difference, we measure the interquartile range (IQR) of the eigenvalues as a summary of how heterogeneous the affordance landscape is across modes. Networks with a wide distribution of costs (that is, some modes that are substantially more expensive than other modes) are considered more heterogeneous compared to networks where the costs are similar across modes. To ensure that our results are not driven by differences in network size, we regress out the number of nodes from the IQR (see Methods for details).

We then assess whether the heterogeneity of the landscape reflects functional specialization. As a measure of function, we utilize the sensorimotor-association (S-A) axis, which ranks brain regions (or collections thereof) according to their involvement in sensorimotor (lower-order) or associative (higher-order) functions~\cite{sydnor2021neurodevelopment}. Strikingly, we find that the landscape heterogeneity of a network correlates with its position along the S-A axis $(r_{S} = - 0.64,\, p = 0.006)$, such that sensory networks have a more heterogeneous landscape, while association networks have a more homogeneous landscape (Fig.~\ref{fig:Fig4}\,\textbf{c}, using the 17-network subdivision of the cortex~\cite{yeo2011organization} for increased statistical power). This relationship suggests that the structural wiring of sensory networks is configured to support a specific set of computations, while association networks maintain more uniform access across a broader range of activity states. The heterogeneous landscape of sensory networks (such as the visual network) might facilitate the filtering and selective excitation of specific modes, enabling these networks to process and refine sensory information~\cite{mesulam1998sensation}. In contrast, the relatively homogeneous landscape of association networks (such as the default mode network) might allow for the exploration of a greater diversity of states, consistent with their role in flexible, domain-general, higher-order, and multimodal processing~\cite{margulies2016situating,vazquez2019gradients,mesulam1998sensation}.

We also examine the spatial structure of the modes in each network; see Fig.~\ref{fig:Fig4}\,\textbf{d}, where we provide examples of two modes that are low in cost (top) and high in cost (bottom) for the visual network of a representative individual. We hypothesize that bilaterally symmetric patterns of activity are lower in cost than patterns that are spatially localized to one hemisphere. To test this hypothesis, we quantify the bilateral symmetry of each mode by defining a bilateral index, which measures how evenly a mode is distributed across the two hemispheres (Eq.~\eqref{eq:bilateral_index}; see Supplementary Information Sec.~\ref{sec:bilateral}). We find that modes with a higher bilateral index are consistently lower in cost across all seven networks (Fig.\,S~\ref{fig:Fig8}, median Spearman $\rho$ ranging from $-0.23$ to $-0.46$ for each network; Wilcoxon signed-rank test, $p < 10^{-16}$ for all networks, $n = 100$ subjects per network). This finding suggests that the structural wiring of cortical networks naturally favors bilaterally symmetric patterns of activity, consistent with the prevalence of bilateral activation observed in the neuroimaging literature~\cite{salvador2005neurophysiological,stark2008regional,toro2008functional}.

\noindent{\large \myfont Affordance landscape heterogeneity increases over training in artificial networks}

\begin{figure}[h!]
    \centering
    \includegraphics[width=0.95\linewidth]{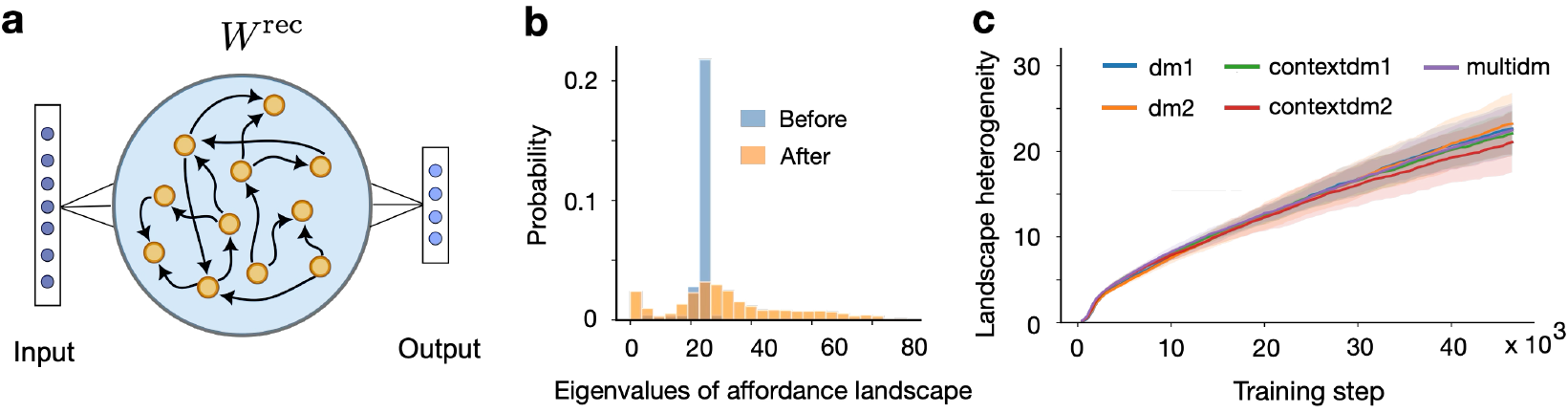} \\
    \raggedright
    \captionsetup{labelformat=empty}
    {\spacing{1.25} \caption{\small \textbf{Fig.~\ref{fig:Fig5} $|$ Affordance landscape heterogeneity increases over training in artificial networks.} 
    (a) Schematic of an artificial recurrent neural network, consisting of an input layer, a recurrent network with weight matrix $W^{\textrm{rec}}$, and an output layer. (b) Distribution of eigenvalues of the affordance landscape of $W^{\textrm{rec}}$ before and after training on a single decision-making task. (c) Heterogeneity of the affordance landscape (IQR of the eigenvalues) as a function of training step, for networks trained on one of five decision-making task variants (dm1, dm2, contextdm1, contextdm2, multidm). Curves indicate the mean over 10 independent seeds and the shaded regions indicate the standard deviation.
\label{fig:Fig5}}}
\end{figure}

Our analysis of neural systems, both at the microscale (ring attractor networks) and at the macroscale (human connectomes), demonstrates how the affordance landscape provides insight into the computations and functions supported by these networks. We now ask a complementary question: how does training for a specific function reshape the affordance landscape? We hypothesize that as a network is trained to perform a single task, we will observe increasing heterogeneity of the affordance landscape. To isolate the effects of training for a single function, we examine artificial neural networks trained on a single task.

We train artificial recurrent neural networks (RNNs) with recurrent weight matrix $W^{\textrm{rec}}$ (Fig.~\ref{fig:Fig5}\,\textbf{a}) consisting of $N = 64$ units on a set of decision-making tasks~\cite{yang2019task} (Methods). We track the affordance landscape of $W^{\textrm{rec}}$ at each training step. Before training, the eigenvalues of the affordance landscape are tightly clustered, indicating that all modes are similar in cost (Fig.~\ref{fig:Fig5}\,\textbf{b}, left). After training, the distribution broadens markedly, with certain modes being low in cost, while other modes become higher in cost (Fig.~\ref{fig:Fig5}\,\textbf{b}, right). Examining the IQR of the eigenvalues over training, we find that this heterogeneity increases monotonically across all five decision-making tasks examined (Fig.~\ref{fig:Fig5}\,\textbf{c}). Training thus actively reorganizes the affordance landscape, and partitions modes into ones that are low in cost and ones that are high in cost. Goal-directed learning can therefore be viewed as sculpting the landscape of affordable computations.

\noindent{\large \myfont Discussion}




The brain supports a wide range of functions through an intricate web of interactions between neural elements~\cite{park2013structural,kulkarni2025toward,bassett2017network,vyas2020computation,bullmore2009complex,power2011functional}. Recent years have seen significant advances in measuring both brain structure~\cite{lin2024network,varshney2011structural,van2013wu} and activity~\cite{urai2022large,demas2021high,cunningham2014dimensionality} with unprecedented precision. Yet, developing a general understanding how the structure of interactions constrains which activity patterns a circuit can readily support, and how that relates to function remains an open area of research~\cite{fotiadis2024structure,honey2010can}. Here, we introduce a general framework that makes these questions concrete by casting computation as a controlled transition of network activity toward a functionally relevant state~\cite{vyas2020computation,gu2015controllability,kim2018role}. This enables us to define a computational affordance landscape, in terms of network structure, that encodes which activity transformations are favorable and which ones are not. Applying this framework to a model of the fly head direction circuit~\cite{noorman2024maintaining,seelig2015neural}, we find that the cheapest computation of the network corresponds precisely to its known function, and that the optimal inputs required to implement it match the known anatomy~\cite{turner2020neuroanatomical,green2017neural}. At the macroscale, affordance landscape heterogeneity varies systematically along the sensorimotor-association axis of the human cortex~\cite{sydnor2021neurodevelopment,margulies2016situating}, reflecting the distinct computational roles of sensory and association networks. Finally, training artificial recurrent neural networks progressively increases landscape heterogeneity, suggesting that learning sculpts the distribution of affordable computations. Thus, although structure-function relationships in the brain are nuanced~\cite{fotiadis2024structure}, our results demonstrate that the affordance landscape provides a tractable and interpretable window into how network structure shapes computational function.

Our framework opens the door to studying how affordance landscapes are actively reshaped in biological systems. Neuromodulators (such as dopamine) alter synaptic interactions and thereby shift the computational properties of neural circuits~\cite{marder2012neuromodulation,fries2005mechanism,reynolds2004attentional,fotiadis2026biological}; within our framework, such modulation can be viewed as a direct means of reshaping the affordance landscape, thereby shifting which computations are low in cost. Similarly, attention is known to modulate effective connectivity between brain regions~\cite{womelsdorf2007modulation,fries2005mechanism}, and our framework provides a quantitative lens through which to study how attentional modulation gates neural computation. Beyond neural systems, the framework applies broadly to any networked system in which function emerges from structure, including physical networks that learn and adapt~\cite{stern2025physical,stern2023learning}. More generally, by making explicit how structure shapes the landscape of affordable computations, our results open the door to the principled design of networks with desired computational properties.

The results presented here are obtained by examining network dynamics in a linear regime, where the affordance landscape can be expressed analytically in terms of network structure. This linearization enables a direct and mechanistic link between structure and computation, with the landscape capturing the network's computational properties in the neighborhood of a given fixed point. Importantly, the framework can be applied to nonlinear systems by linearizing about a fixed point, as we demonstrate for the ring attractor network. Here, the rotational symmetry of the circuit ensures that the linearization is similar at every stable bump, such that examining the affordance landscape around a single bump enabled us to characterize the computational properties of the full circuit. More generally though, the framework can be extended to arbitrary operating points by estimating the Jacobian either analytically or directly from data~\cite{eisen2026characterizing}, allowing the affordance landscape to be computed locally across the full nonlinear dynamics. A second practical consideration is that computing the affordance landscape requires an eigendecomposition of the Gramian, which can become costly for large connectomes. Future work can address this through approximation schemes, for example by exploiting low-rank structure in the Gramian or employing scalable numerical methods. Together, these extensions would broaden the applicability of the framework while preserving its interpretability.

\newpage

\noindent {\large \myfont \textbf{Methods}}
\vspace{-28pt}

\noindent\rule{\textwidth}{.5pt}

\begin{methods}
\setlength{\parindent}{0pt}

\subsection{Network control theory} 
We consider a network of $N$ nodes and associate with each node a scalar that captures its activity. We collect the activity of all nodes into a vector $\mathbf{x}(t) \in \mathbb{R}^N$, which we refer to as the state of the network. The state evolves through a general function $f$ such that $\dot{\mathbf{x}}(t) = f(t, \mathbf{x}(t) \,, \mathbf{u}(t))$, where $\mathbf{u}(t) \in \mathbb{R}^m$ are the inputs to the network. For analytical tractability, we linearize the dynamics about an equilibrium point ($\mathbf{x}^*$, $\mathbf{u}^*$), and we express the dynamics of the deviations $\mathbf{r}(t) = \mathbf{x}(t) \,-\, \mathbf{x}^*$ as $\dot{\mathbf{r}}(t) = W \,\mathbf{r}(t) \,+\, B \, \mathbf{v}(t)$, where $W$ is the Jacobian of $f$, $B$ captures the structure of inputs to the network, and $\mathbf{v}(t) = \mathbf{u}(t) - \mathbf{u}^*$ are the deviations in the input about equilibrium. The state of the network at any time $T$ is \begin{equation}
    \mathbf{r}(T) = e^{WT} \mathbf{r}_0 + \int_0^T e^{Ws} B \, \mathbf{v}(s) \, ds,
\end{equation}
where $\mathbf{r}_0$ is the initial state of the network. To drive the network from an initial state $\mathbf{r}(0) = \mathbf{r}_0$ to a target state $\mathbf{r}(T) = \mathbf{r}_f$, the inputs must compensate for the discrepancies $\mathbf{D} = \mathbf{r}_f \, - \, e^{WT} \mathbf{r}_0$ between the target state $\mathbf{r}_f$ and the homogeneous solution $e^{WT} \mathbf{r}_0$ (i.e., the state of the network at time $T$ for $\mathbf{v}(t) = 0$). We quantify the cost of driving the network by the energy of the input, $\mathcal{C} = \int_0^T \| \mathbf{v}(t)\|^2 \, dt$. The input that can achieve this target state with the least cost is:~\cite{kim2020linear, parkes2024network, bechhoefer2021control}
\begin{equation}\label{eq:optimal_input}
    \mathbf{v}^{\mathrm{min}}(t) = B^{\top} e^{W^{\top}(T - t)} \, \mathcal{W}^{-1} \, \mathbf{D} \quad \mathrm{where} \quad \mathcal{W} = \int_0^T e^{Ws} B B^{\top} e^{W^{\top}s} \, ds.
\end{equation}
The resulting minimum cost is $\mathcal{C}^{\mathrm{min}} = \mathbf{D}^{\top} \mathcal{W}^{-1} \mathbf{D}$. Since the inverse Gramian $\mathcal{W}^{-1}$ shapes the cost of driving the network to any target state, we refer to it as the computational affordance landscape. 

\subsection{Computing the affordance landscape} 
The matrix $\mathcal{W}^{-1}$ [see Eq.~\ref{eq:optimal_input}] defines the computational landscape. Throughout, we set $T = 1$ and we assume that all nodes receive inputs independently, so that $B = \mathbb{I}$. For a general network, we compute $\mathcal{W}$ by numerical integration, then invert it numerically and compute its eigendecomposition. For symmetric networks $(W = W^{\top})$, the eigenvalues of the affordance landscape admit a closed form $w_i^{-1}  = 2 \lambda_i/(e^{2\lambda_i T} - 1)$, where $\lambda_i$ are the eigenvalues of $W$.

\subsection{Ring attractor network} 
The ring attractor network consists of $N$ neurons that are arranged topologically in a ring, with preferred orientations that uniformly tile $(0, 2\pi]$ with spacing $\Delta \theta = 2\pi/N$ between consecutive neurons. The connectivity consists of local excitation between similarly tuned neurons and global inhibition, such that $W_{jk} = J_I \, + \, J_E \, \cos (\theta_j - \theta_k)$. The total input activity to each neuron $h_j$ evolves according to
\begin{equation}\label{eq:ring_attractor_eq_methods}
    \tau \frac{d h_j(t)}{dt} = - h_j(t) + \frac{1}{N} \sum_k W_{jk} \, \Phi(h_k) + I_j(t),
\end{equation}
where $\Phi(\cdot) = (1 \,+\, \tanh(x))/2$ is a nonlinear transfer function mapping the input to the firing rate, $I_j(t)$ is any external input to the neuron, and $\tau$ is the neural time constant. We simulate a network of $N = 25$ neurons with $(J_I, J_E) = (-9,10)$ and $\tau = 0.01$s. We add Gaussian noise to the weight matrix by drawing each entry independently from $\mathcal{N}(0, \sigma)$ with $\sigma = 0.5$ and we symmetrize the weight matrix by considering $(W \, + \, W^{\top})/2$. We compute the affordance landscape by linearizing Eq.~\eqref{eq:ring_attractor_eq} about a stable bump $\mathbf{h}^*$ (see Supplementary Information Sec.~\ref{sec:SI_ring_attractor}). The optimal input for rotating the bump is computed analytically using Eq.~\eqref{eq:optimal_input}.

\subsection{Human Connectome Project (HCP) data.}

We derive structural connectivity matrices from the Human Connectome Project (HCP) ~\cite{van2013wu}. The HCP study was approved by the Washington University Institutional Review Board and informed consent was obtained from all subjects. We analyze diffusion magnetic resonance imaging data from a sample of 100 unrelated healthy subjects (54\% female; mean age $= 29.1 \pm 3.7$ years; age range $= 22$--$36$ years), first minimally preprocessed using the HCP consortium pipelines (https://github.com/Washington-University/HCPpipelines), including correction for imaging distortions, subject motion, gradient nonlinearities, and registration to each subject's native anatomical space~\cite{glasser2013minimal}. We then construct structural connectomes in MRtrix3~\cite{tournier2019mrtrix3} using multi-shell, multi-tissue constrained spherical deconvolution, anatomically-constrained probabilistic tractography, and SIFT2 weighting~\cite{smith2015sift2} to improve biological accuracy and reduce known tractography biases~\cite{tournier2010improved}, as previously described~\cite{fotiadis2023myelination}. We generate subject-specific symmetric, weighted structural connectivity matrices based on the Schaefer parcellation~\cite{schaefer2017local} ($400 \times 400$ brain regions), with edge weights defined as the SIFT2-weighted streamline count normalized by the gray matter volumes of the connected regions. We assign each of the 400 regions to one of 7 or 17 functional networks using the Yeo atlas~\cite{yeo2011organization}. Left and right hemisphere assignments are determined from the Schaefer-400 parcel labels. For each network, we compute the affordance landscape. We summarize each landscape by the interquartile range (IQR) of its eigenvalues as a measure of heterogeneity. Since networks differ in size, we use linear regression to regress out the number of nodes from the IQR before comparing across networks. Network positions along the sensorimotor-association axis are taken from Sydnor et al.~\cite{sydnor2021neurodevelopment}.

\subsection{Artificial recurrent neural networks}
We train recurrent neural networks (RNN) individually on five perceptual decision-making tasks: DM1, DM2, ContextDM1, ContextDM2, and MultSen DM, following Yang et al.~\cite{yang2019task} and the implementation of Gu et al~\cite{gu2025task}. The network activity $\mathbf{r}(t)$ evolves according to
\begin{equation}
    \tau \frac{d \mathbf{r}}{dt} = - \mathbf{r} + f(W_{\textrm{rec}} \,+\, W_{\textrm{in}} \mathbf{u} \,+\, \mathbf{b} \,+\, \sqrt{2 \tau \sigma^2_{\textrm{rec}}}\epsilon), 
\end{equation}
where $\tau = 100$ ms is the time constant, $W_\textrm{rec} \in \mathbb{R}^{N \times N}$ is the recurrent weight matrix, $W_\textrm{in}$ is the input weight matrix, $\mathbf{u}(t)$ is the input to the network, $f(\cdot) = \tanh$ is the nonlinearity, and $\epsilon$ is Gaussian white noise. We consider networks that comprise 64 recurrent units. For each task, the network received inputs of three types: a fixation signal, stimulus inputs, and a one-hot rule signal identifying the current task. For further details on the task setup and network architecture, see Yang et al.~\cite{yang2019task}. For each task, we train networks independently across 10 random seeds for $3 \times 10^6$ trials, and we save checkpoints of the model every 500 gradient steps to obtain snapshots of $W_\textrm{rec}$ across the full training trajectory. We compute the affordance landscape of $W_{\textrm{rec}}$ at each checkpoint.

\end{methods}

\section*{Data Availability}

The data analyzed in this paper are openly available at: \\
\href{https://github.com/SumanSKulkarni/NetworkComputation}{https://github.com/SumanSKulkarni/NetworkComputation}

\section*{Code Availability}

The code used to perform the analyses in this paper is openly available at:\\
\href{https://github.com/SumanSKulkarni/NetworkComputation}{https://github.com/SumanSKulkarni/NetworkComputation}

\begin{addendum}

\item[Supplementary Information.] Supplementary text and figures accompany this paper.

\item[Acknowledgments.] We thank Valerie J. Sydnor, Vivek Jayaraman, Ann M. Hermundstad, Shi Gu, and Yuhang Wu for helpful conversations. We also thank Isabella Stallworthy, Leo Chambers, and Kat Hefter for both useful discussions and comments on earlier versions of this manuscript. The authors acknowledge support from the Army Research Office MURI program [W911NF2410228]. The content is solely the responsibility of the authors and does not necessarily represent the official views of any of the funding agencies.
 
\item[Competing Interests.] The authors declare no competing financial interests.
 
\item[Corresponding Author.] Correspondence and requests for materials should be addressed to D.S.B. \\(dani.bassett@yale.edu).
 
\end{addendum}







\setcounter{figure}{0}

\noindent {\large \myfont \textbf{Supplementary Materials}}

\section{Cost of target oscillations}\label{sec:oscillations}

We consider a network of $N$ nodes whose firing rates $\mathbf{r}(t)$ about a stable operating point evolve according to the rate equation
\begin{equation}\label{eq:oscillation_dynamics}
    \frac{d \mathbf{r}(t)}{dt} = W \mathbf{r}(t) + B \, \mathbf{u}(t),
\end{equation}
where $W \in \mathbb{R}^{N \times N}$ is the weighted connectivity matrix, $\mathbf{u}(t) \in \mathbb{R}^{m}$ are inputs to the network, and $B \in \mathbb{R}^{N \times m}$ is the input matrix that routes the inputs to the nodes. The output of the network is $\mathbf{y}(t) = C \, \mathbf{r}(t)$, where $C \in \mathbb{R}^{k \times N}$ is a readout matrix. We define our target as a sustained oscillatory output at the readout 
\begin{equation}
    \mathbf{y}^{*}(t) = \mathrm{Re}\!\left\{\mathbf{y}_f\, e^{i\omega t}\right\},
\end{equation}
where $\mathbf{y}_f \in \mathbb{C}^k$ encodes the target amplitude and phase at each readout node and $\omega$ is the target frequency. The cost of producing these target oscillations is the input energy expended over one period $\tau = 2\pi/\omega$
\begin{equation}\label{eq:oscillation_cost}
\mathcal{C} = \int_0^{\tau} \|\mathbf{u}(t)\|^2 \, dt.
\end{equation}

We now derive the minimum cost of producing these oscillations. Since the dynamics in Eq.~\eqref{eq:oscillation_dynamics} are linear, oscillatory inputs at frequency $\omega$ produce oscillatory outputs at the same frequency; any other input waveform would excite additional frequency components that increase the cost without contributing to the target. We therefore restrict to inputs of the form $\mathbf{u}(t) = \mathrm{Re} \left\{\mathbf{u}_0\, e^{i\omega t}\right\}$, where $u_0 \in \mathbb{C}^m$ specifies the amplitude and phase at each input node. For inputs of this form, the cost simplifies to $\mathcal{C} = \frac{\pi}{\omega}||\mathbf{u}_0\|^2$, such that minimizing the cost reduces to finding the minimum-norm $\mathbf{u}_0$ that produces the target oscillation.

To find which $\mathbf{u}_0$ produces the target, we solve the network equation. The general solution to Eq.~\eqref{eq:oscillation_dynamics} consists of a homogeneous solution and a particular solution. Substituting a particular solution of the form $\mathbf{r}_{p}(t) = \mathrm{Re}\left\{\mathbf{v}\, e^{i\omega t}\right\}$ into Eq.~\eqref{eq:oscillation_dynamics} yields $\mathbf{v} = (i\omega I - W)^{-1} \, B \, \mathbf{u}_0$. The general solution is therefore
\begin{equation}
    \mathbf{r}(t) = \underbrace{\mathbf{c} \, e^{Wt} \vphantom{(i\omega I - W)^{-1}}}_{\mathrm{homogeneous}} \, +\, \underbrace{(i\omega I - W)^{-1} \, B \, \mathbf{u}_0 \, e^{i\omega t}}_{\mathrm{particular}},
\end{equation}
where $\mathbf{c}$ is a constant that depends on the initial conditions. Since $W$ is stable, the homogeneous part decays to $0$ as $t \to \infty$ and the steady-state output is
\begin{equation}\label{eq:steady_state_transfer_function}
    \mathbf{y}_\mathrm{ss}(t) = \mathrm{Re}\!\left\{H(i\omega)\,\mathbf{u}_0\, 
e^{i\omega t}\right\}, \quad \textrm{where} \quad   H(i\omega) = C\,(i\omega I - W)^{-1} B \;\in\; \mathbb{C}^{k \times m}
\end{equation}
is the transfer function. For the steady-state output $\mathbf{y}_\mathrm{ss}$ to match the target $\mathbf{y}^*$, we require $H(i\omega)\,\mathbf{u}_0 = 
\mathbf{y}_f$.
Minimizing the cost subject to 
$H(i\omega)\,\mathbf{u}_0 = \mathbf{y}_f$ results in the solution
\begin{equation}\label{eq:oscillatory_cost}
\mathbf{u}_0^* = H(i\omega)^\dagger \left(H(i\omega)\,H(i\omega)^\dagger 
\right)^{-1} \mathbf{y}_f \quad \textrm{such that} \quad
\mathcal{C}^{\mathrm{min}} = \frac{\pi}{\omega}\;\mathbf{y}_f^*\;\mathcal{W}(\omega)^{-1}\;\mathbf{y}_f,
\end{equation}
where
\begin{equation}\label{eq:oscillatory_gramian}
    \mathcal{W}(\omega) = H(i\omega)\,H(i\omega)^\dagger \;\in\; 
    \mathbb{C}^{k \times k}
\end{equation}
is the oscillatory affordance landscape. Like the Gramian $\mathcal{W}$ for target states, $\mathcal{W}(\omega)$ encodes how network structure $W$, input architecture $B$, and readout structure $C$ jointly determine which oscillatory targets $\mathbf{y}_f$ are intrinsically cheap or costly to produce at each frequency $\omega$. 

Our framing of computations enables us to build a mechanistic account of how the cost of a computation depends on network structure through the transfer function. Below, we use this account to study two phenomena observed empirically in neural circuits: frequency-selective filtering and frequency-dependent selective routing.

\subsection{Frequency-selective filtering in neural circuits\\}


Neural circuits respond preferentially to oscillatory input at specific frequencies, acting as filters whose selectivity is shaped by circuit-level properties such as synaptic interactions~\cite{ito2007frequency,campbell2025hardwired}. This selectivity is also shown to support neural computations; for example, circuit connectivity can give rise to a linear filter that computes temporal difference errors for reward-based learning~\cite{campbell2025hardwired}. Here we provide a structural and mechanistic account of how interactions in a circuit give rise to specific responses to oscillatory inputs.

For intuition, we consider a simple setting where a single input node $i$ receives an input $u(t)$ (so that $B = \mathbf{e}_i$, where $\mathbf{e}_i \in \mathbb{R}^N$ is the standard basis vector) and we track the activity of a single output node $j$ (so that $C = \mathbf{e}_j^\top$). The transfer function [see Eq.~\ref{eq:steady_state_transfer_function}] in this setting is a scalar
\begin{equation}
    H(i \omega) = \mathbf{e}_j^{\top} (i\omega I - W)^{-1} \,\mathbf{e}_i.
\end{equation}
The cost to produce oscillations of frequency $\omega$ and amplitude $y_f$ at node $j$ is
\begin{equation}
      \mathcal{C}^{\mathrm{min}} (\omega) = \frac{\pi}{\omega} \frac{|y_f|^2}{|H(i \omega)|^2} . 
\end{equation}
The cost depends on frequency through the transfer function and is lowest at a frequency $\omega^*$ where $|H(i \omega)|$ is largest. Oscillations at frequencies away from $\omega^*$ are substantially more costly.

To illustrate how this frequency-selectivity depends on network structure, we consider two networks with distinct network structure $W$ and plot the cost of producing oscillations of unit amplitude (that is, $|y_f| = 1$) as function of $\omega$ in Fig.~\ref{fig:Fig2}(c). The cost curves differ markedly across the two networks, with minima at different frequencies, reflecting how network structure shapes the oscillatory affordance landscape. Moreover, we also note that neuromodulators such as dopamine can change the strength of interactions (that is, $W$), thereby shifting $\omega^{\textrm{min}}$ and changing which oscillatory computations the circuit supports at low cost~\cite{ito2007frequency}.

\begin{figure}[h!]
    \centering
    \includegraphics[width=\linewidth]{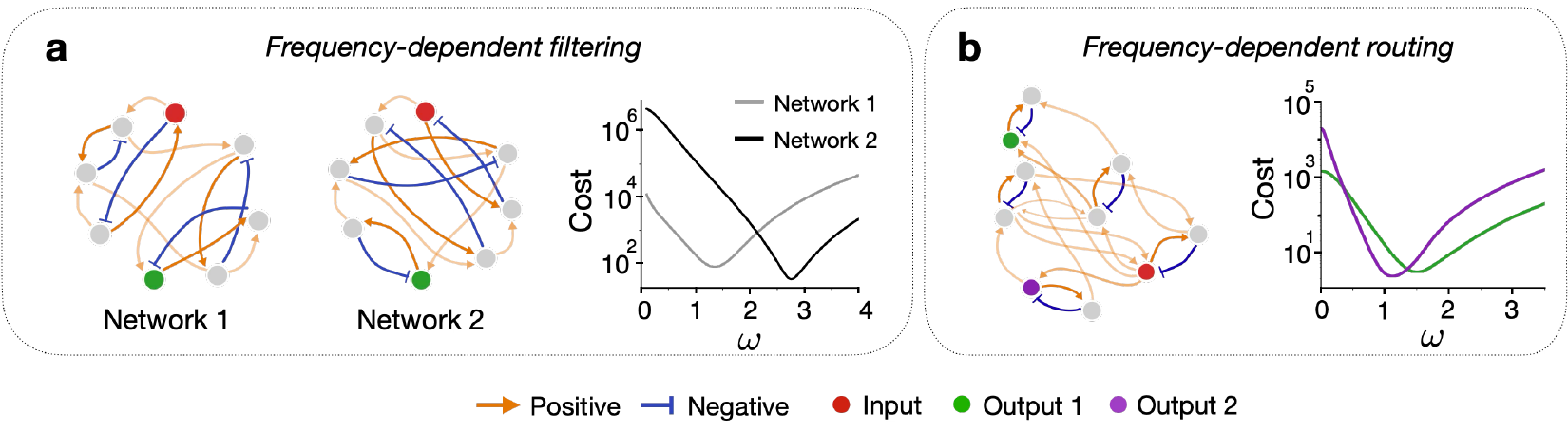} \\
    \raggedright
    \captionsetup{labelformat=empty}
    {\spacing{1.25} \caption{\small \textbf{Fig.\,S~\ref{fig:Fig6} $|$ Cost of producing target oscillations in a network} \textbf{a}, \textit{Frequency-dependent filtering.} (Left) Two networks with the same input and output, but distinct network structure. (Right) Cost of producing a unit-amplitude oscillation at the output node, as a function of oscillation frequency $\omega$ for each network. \textbf{b}, \textit{Frequency-dependent routing.} (Left) A single network with one input and two output nodes. (Right) Cost of producing a unit-amplitude oscillation at each output node as a function of oscillation frequency $\omega$. \label{fig:Fig6}}}
\end{figure}

\subsection{Frequency-dependent routing in neural circuits \\}

Network structure shapes the transfer function differently at different nodes, such that distinct parts of a circuit respond preferentially to different frequencies, effectively creating distinct communication channels. Empirically, neural circuits have been shown to use oscillatory frequency as a channel for selectively routing information, such that distinct frequencies carry information along distinct pathways~\cite{colgin2009frequency, khamechian2019routing}.

To illustrate, we consider a network with a single input node $i$ and two output nodes $j_1$ and $j_2$. The transfer function from input node $i$ to output node $j \in \{j_1, j_2\}$ is
\begin{equation}\label{eq:transfer_selective_routing}
    H_j(i\omega) = \mathbf{e}_j^\top (i\omega I - W)^{-1} \mathbf{e}_i = 
    \sum_{\ell=1}^{N} \frac{(\mathbf{e}_j^\top \mathbf{v}_\ell)\,
    (\mathbf{w}_\ell^\top \mathbf{e}_i)}{i\omega - \lambda_\ell},
\end{equation}
where $\mathbf{v}_\ell$ and $\mathbf{w}_\ell$ are the right and left eigenvectors of $W$ with eigenvalue $\lambda_\ell$. The contribution of each mode $\ell$ decomposes into three factors: the output alignment $\mathbf{e}_j^\top \mathbf{v}_\ell$ (how strongly mode $\ell$ is expressed at node $j$), the input alignment $\mathbf{w}_\ell^\top \mathbf{e}_i$ (how strongly the input at node $i$ excites mode $\ell$), and a resonance factor $(i\omega -\lambda_\ell)^{-1}$ that amplifies contributions near the natural frequency $\mathrm{Im}(\lambda_\ell)$ of the mode.

Because nodes $j_1$ and $j_2$ weight the eigenmodes differently through their output alignments, $H_{j_1}(i\omega)$ and $H_{j_2}(i\omega)$ peak at distinct preferred frequencies $\omega_{j_1}^*$ and $\omega_{j_2}^*$, giving rise to distinct oscillatory affordance landscapes $\mathcal{W}_j(\omega) = |H_j(i\omega)|^2$ at each node. In Fig.~\ref{fig:Fig2}(d), we plot the cost of producing oscillations of unit amplitude as a function of frequency $\omega$ at the two nodes. The cost curves have distinct minima: node $j_1$ sustains oscillations at $\omega_{j_1}^{\textrm{min}}$ at low cost, while node $j_2$ sustains oscillations at $\omega_{j_2}^{\textrm{min}}$ at low cost. A single network, driven by a common input, thus supports distinct oscillatory computations at different circuit locations. Network structure determines not only which frequencies are supported, but where in the circuit they are expressed, offering a structural basis for the communication-through-coherence hypothesis, whereby oscillatory synchrony between populations selectively gates 
information flow~\cite{fries2015rhythms}.

\section{Ring attractor dynamics}\label{sec:SI_ring_attractor}

We consider a network of $N$ neurons that are ordered according to their preferred heading direction $\theta_j$ [Fig.~\ref{fig:Fig3}\,\textbf{a}]. We take these preferred heading directions to be evenly spaced by $\Delta \theta = 2 \pi/N$ radians. The neurons are connected to one another recurrently according to a weight matrix whose entries are $W_{jk} = J_I \,+ \, J_E \, \cos(\theta_j - \theta_k)$, where $J_E$ and $J_I$ are the strengths of the local excitation and global inhibition [see Fig.~\ref{fig:Fig3}\,\textbf{b} for the resulting weight matrix]. Each neuron transforms its inputs into firing rates via a nonlinear activation function $\Phi(\cdot)$. The total input $h_j$ to each neuron is governed by
\begin{equation}\label{eq:ring_attractor_eq_SI}
    \tau \frac{d h_j(t)}{dt} = - h_j(t) + \frac{1}{N} \sum_k W_{jk} \, \Phi(h_k) + I_j(t),
\end{equation}
where $I_j(t)$ captures any external input to the neuron (i.e., from other sources outside the ring of $N$ neurons) and $\tau$ is a neural time constant. We take $\Phi(x) = [1 \,+\, \tanh(x)]/2$. 

For appropriate values of local excitation and broad inhibition $(J_E, J_I)$, the network generates a continuum of localized bumps of activity that are stable at any heading $\theta \in [0, 2\pi)$~\cite{noorman2024maintaining}, as shown in Fig.~\ref{fig:Fig3}(c). In the absence of any external inputs, the fixed-point equation is
\begin{equation}\label{eq:fixed_point_ring}
    h_j^* = \frac{1}{N} \sum_{k} W_{jk} \, \Phi(h_k^*).
\end{equation}

To study the affordance landscape about a stable bump, we linearize the dynamics in Eq.~\ref{eq:ring_attractor_eq_SI} about a stable bump $\mathbf{h^*}$ at some orientation $\theta^*$. The linearized dynamics are
\begin{equation}\label{eq:linearized_ring}
    \tau \frac{d\mathbf{y}(t)}{dt} \approx \left[ \mathbb{I} - \frac{1}{N} W \, \mathrm{diag}\left(\Phi^{\prime}(\mathbf{h}^*)\right) \right] \mathbf{y}(t) \, + \, \mathbb{I} \, \mathbf{u}(t), 
\end{equation}
where $\mathbf{y}(t) = \mathbf{h} \, - \, \mathbf{h}^*$ is the deviation from the fixed point activity, $\mathbf{u}(t)$ is the external input to the network, $\mathrm{diag}$ is a diagonal matrix with entries $[\Phi^{\prime}(\mathbf{h}^*)]_{k} = \mathrm{sech}^2(h_k^*)/2$, and $\mathbb{I}$ is the identity matrix. The linearized dynamics can be expressed as 
\begin{equation}
    \tau \, \dot{\mathbf{y}}(t)= J\, \mathbf{y}(t) \,+\, \mathbb{I} \, \mathbf{u}(t), \quad \mathrm{where} \quad J = \mathbb{I} - \frac{1}{N} W \, \mathrm{diag}\left(\Phi^{\prime}(\mathbf{h}^*)\right) 
\end{equation}
is the Jacobian matrix. The eigenvalues of the affordance landscape are then given by
\begin{equation}\label{eq:eig_cost}
    w_i^{-1} = \frac{2\lambda_i}{e^{2 \lambda_iT} - 1},
\end{equation}
where $\lambda_i$ are the eigenvalues of the Jacobian $J$ and $T$ is the time horizon over which the computation occurs. Since the fixed points are marginally stable, the eigenvalues of $J$ satisfy $\lambda_i(J) \leq 0$. As a consequence of the eigenvalues of the affordance landscape being decreasing functions of $\lambda_i$ [See Eq.~\eqref{eq:min_cost_eigen}], it follows that the lowest cost mode corresponds to $\lambda_i(J) = 0$.

To interpret this finding, we examine the kernel of the Jacobian. We note that the stable configurations $\mathbf{h}^*$ are parameterized by $\theta$, and we differentiate Eq.~\eqref{eq:fixed_point_ring} with respect to $\theta$ to obtain
\begin{align}\label{eq:kernel_rotation}
    &\frac{\partial h_j^*}{\partial \theta} = \frac{1}{N} \sum_k W_{jk} \frac{\partial \, \Phi(h^*_k)}{\partial \, h^*_k} \frac{\partial \, h^*_k}{\partial \, \theta} \notag \\ 
    \implies &\underbrace{\left[ \, \mathbb{I} - \frac{1}{N} W \, \mathrm{diag} \left( \Phi^{\prime}(\mathbf{h}^*)\right) \,  \right]}_{J} \frac{\partial\,\mathbf{h}^*}{\partial \theta} = 0.
\end{align}
We therefore observe that the derivative of the bump with respect to $\theta$ (i.e., the direction that corresponds to rotating the bump) lies in the Kernel of the Jacobian $J$. Together, this implies that the direction along which computations are cheapest corresponds precisely to rotation of the bump; that is, the network's cheapest afforded operation (from structure and dynamics) is precisely the one needed to update its representation of heading direction (which is its function). 

\begin{figure}[h!]
    \centering
    \includegraphics[width=0.65\linewidth]{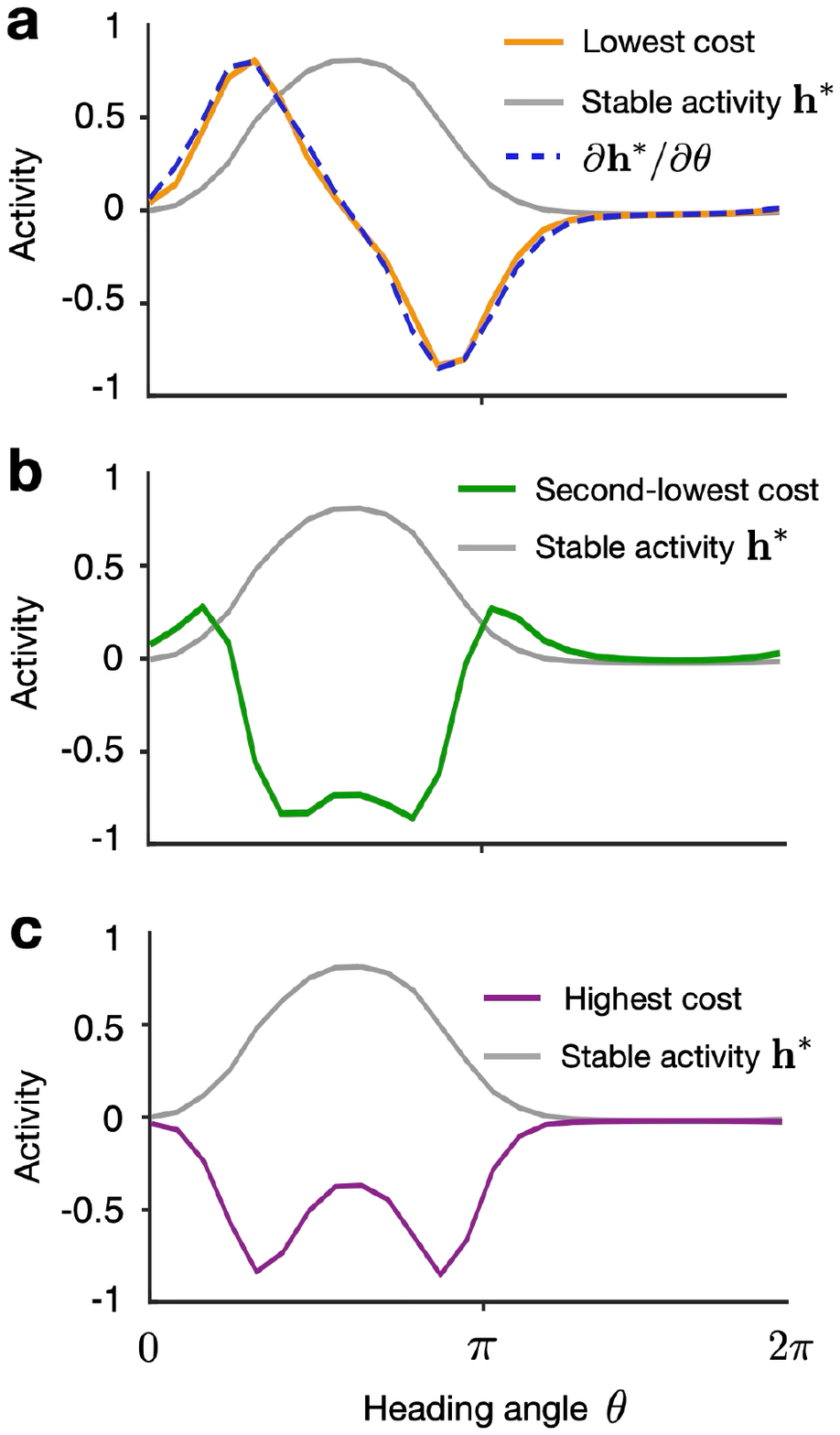} \\
    \raggedright
    \captionsetup{labelformat=empty}
    {\spacing{1.25} \caption{\small \textbf{Fig\,S~\ref{fig:Fig7} $|$ Modes of the affordance landscape for the ring-attractor network.} In each panel, the stable bump of activity $\mathbf{h}^*$ is shown in gray for reference. \textbf{a}, The lowest-cost mode of the affordance landscape and the derivative of the stable bump with respect to heading, $\partial \mathbf{h}^*/ \partial \theta$. \textbf{b}, The second lowest-cost mode of the affordance landscape. \textbf{c}, The highest-cost mode of the affordance landscape. \label{fig:Fig7}}}
\end{figure}

We numerically examine a ring attractor network with $N = 25$ neurons with a small amount of noise added to the weight matrix (Methods). We then use Eq.~\ref{eq:linearized_ring} to compute the computational affordance landscape (Fig.~\ref{fig:Fig3}\,\textbf{d}). The lowest-cost mode corresponds precisely to the derivative of the bump with respect to heading (Fig.,S~\ref{fig:Fig7},\textbf{a}), consistent with rotation being the cheapest computation. We also show the second-lowest cost mode (Fig.\,S~\ref{fig:Fig7}\,\textbf{b}) and the highest cost mode (Fig.\,S~\ref{fig:Fig7}\,\textbf{c}). While interpreting these other modes requires further investigation, the second lowest-cost mode resembles a broadening or sharpening of the bump (Fig.\,S~\ref{fig:Fig7}\,\textbf{b}), while the highest-cost mode corresponds to a net increase or decrease in the amplitude of the bump (Fig.\,S~\ref{fig:Fig7}\,\textbf{c}). The high cost of the latter seems to be consistent with the function of the ring attractor in maintaining persistent internal representations of heading.

\section{Bilateral patterns of activation}\label{sec:bilateral}

\begin{figure}[h!]
    \centering
    \includegraphics[width=0.95\linewidth]{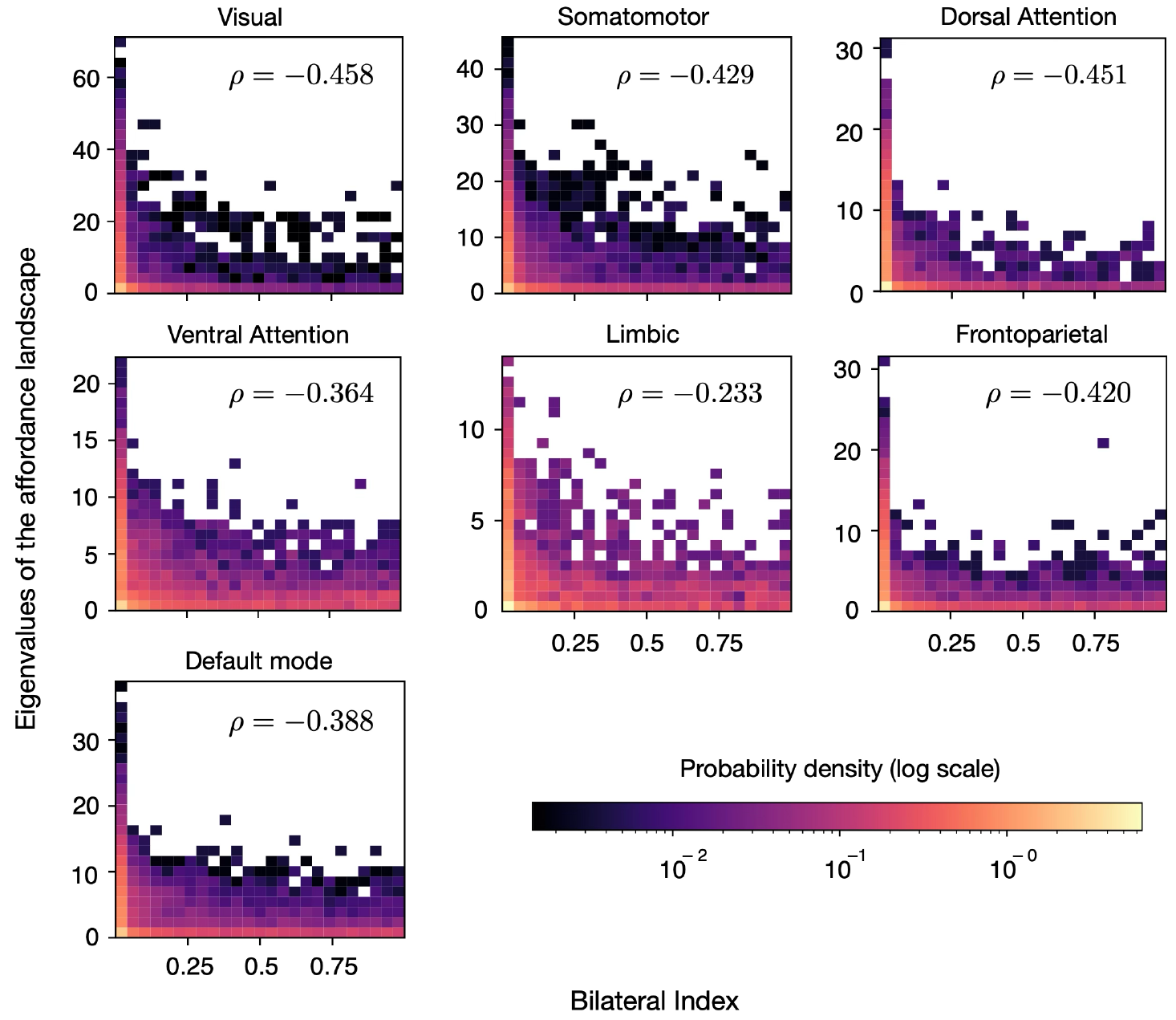} \\
    \raggedright
    \captionsetup{labelformat=empty}
    {\spacing{1.25} \caption{\small \textbf{Fig\,S~\ref{fig:Fig8} $|$ Bilateral patterns of activation are energetically favorable.} Two-dimensional density histograms showing the relationship between the bilateral index (BI) of a mode (x axis) and the cost of driving it (y axis) for each of the seven canonical networks, pooled across all 100 subjects. The bilateral index, defined in Eq.~\eqref{eq:bilateral_index}, quantifies how evenly a mode is distributed across the two hemispheres, ranging from 0 (activity confined to one hemisphere) to 1 (activity balanced equally across both hemispheres). Color indicates probability density (log scale). Each panel reports the median Spearman correlation $\rho$ between the cost of a mode and its bilateral index, computed separately for each subject and summarized across subjects. The negative relationship between cost and bilateral index is consistent and significant across all networks (Wilcoxon signed-rank test, $p<10^{-16}$ for all $7$ networks with $N = 100$ subjects).\label{fig:Fig8}}}
\end{figure}

We test the hypothesis that bilaterally symmetric activity patterns are lower in cost than patterns that are lateralized to one hemisphere. For each mode $\mathbf{v}_i$ of the affordance landscape, we partition the parcels into left- and right-hemisphere sets and we measure the fraction of the mode's amplitude in each hemisphere as $E_L = \sum_{i \in L} |v_i|^2$ and $E_R = \sum_{i \in R} |v_i|^2$, where the subscript denotes the left and right hemisphere, respectively. 
We define the laterality index $\mathrm{LI} = (E_L - E_R)/(E_L + E_R) \in [-1, 1]$, which measures the degree to which a mode is concentrated in one hemisphere. We then define the bilateral index as
\begin{equation}
\mathrm{BI} = 1 - |\mathrm{LI}| \in [0, 1], \label{eq:bilateral_index}
\end{equation}
which ranges from $0$ (mode confined entirely to one hemisphere) to $1$ (mode balanced equally across both hemispheres). We compute the BI for every mode of every network across all 100 subjects from the HCP and examine its relationship to the mode's cost (Fig.\,S~\ref{fig:Fig8}). To assess statistical significance, we compute the Spearman correlation between mode cost and BI separately for each subject and test whether the resulting distribution of per-subject correlations is significantly negative using a Wilcoxon signed-rank test. We find a consistent negative relationship between cost and bilateral index across all seven networks, with median Spearman $\rho$ ranging from $-0.23$ (Limbic) to $-0.46$ (Visual). The relationship is significant in every network (Wilcoxon signed-rank test, $p<10^{-16}$ for all networks, $N = 100$ subjects per network), confirming that bilaterally symmetric modes are lower in cost than lateralized ones and that this effect is robust across individuals.

\newpage

\section*{\large References}
\vspace{-30pt}
\noindent\rule{\textwidth}{.5pt}
\bibliography{kulkarni_network_computation}

\end{document}